\documentclass[aps,prl,showpacs,twocolumn,superscriptaddress]{revtex4}
\usepackage{amsmath, amssymb}
\usepackage{braket}
\usepackage{ulem}
\usepackage{soul}
\usepackage{dcolumn}
\usepackage{bm}
\usepackage{graphicx}
\usepackage{wasysym}
\usepackage{setspace}
\usepackage{mathrsfs}
\usepackage{color}
\usepackage{hyperref}
\usepackage{CJKutf8}
\hypersetup{colorlinks=true,linkcolor=blue,filecolor=blue,citecolor = blue,urlcolor=blue,}

\usepackage{tikz}
\usepackage{lipsum}

\begin{document}
\title{Chiral Damping-Induced Chirality Switching and Control of Domain Walls in Antiferromagnets}
\author{Collins Ashu Akosa}
\email{collins.akosa@aoni.waseda.jp}
\affiliation
{Department of Applied Physics, Waseda University, Okubo, Shinjuku-ku, Tokyo 169-8555, Japan}
\affiliation{Department of Theoretical and Applied Physics, African University of Science and Technology (AUST), Km 10 Airport Road, Galadimawa, Abuja F.C.T, Nigeria}
\author{Mu-Kun Lee}
\email{mukunlee01@gmail.com}
\affiliation
{Department of Applied Physics, Waseda University, Okubo, Shinjuku-ku, Tokyo 169-8555, Japan}
\author{Aur\'elien Manchon}
\affiliation{Aix-Marseille Universit\'e, CNRS, CINaM, Marseille, France}
\author{Masahito Mochizuki}
\affiliation
{Department of Applied Physics, Waseda University, Okubo, Shinjuku-ku, Tokyo 169-8555, Japan}
\date{\today}
\begin{abstract}  

We investigate the impact of chiral damping (CD) on current-driven domain-wall (DW) dynamics in antiferromagnets (AFMs). 
Asymmetric CD between sublattices generates off-diagonal components in the DW mass tensor, thereby coupling translational and rotational modes. When CD is modulated by an ac gate voltage via the Rashba spin-orbit interaction (RSOI), symmetric and asymmetric contributions induce oscillations in the DW velocity and tilt angle, respectively. A perturbative analysis yields explicit expressions for the oscillation amplitudes, in quantitative agreement with numerical simulations. Remarkably, even in the absence of Dzyaloshinskii–Moriya interaction (DMI), asymmetric CD enables chirality switching between N\'eel- and Bloch-type DWs. Finally, by exploiting the relativistic Lorentz contraction of the DW width at high driving currents, we propose an experimentally viable protocol to qualitatively and quantitatively extract the CD contribution. These results establish clear experimental signatures of CD in antiferromagnetic DW dynamics and demonstrate its potential as a control parameter for magnetic textures.

\end{abstract}
\maketitle
\textit{Introduction}.---The quest for faster, smaller, and more durable spintronic devices has fueled intensive research into magnetic systems that combine energy efficiency with robustness \cite{Allwood2005,Chappert2007,Parkin2008}. Conventional ferromagnetic (FM) platforms, despite their maturity, face fundamental limitations: dipolar interactions give rise to stray-field effects that hinder miniaturization and complicate device integration \cite{Krishnia2014,Murapaka2014,Obrien2009}. AFMs provide a compelling alternative. Their compensated spin structure eliminates stray fields, they support ultrafast spin dynamics, and they enable efficient spin manipulation, positioning AFMs as promising building blocks for next-generation memory technologies \cite{Gomonay2014,Wadley2016,Jungwirth2016,Baltz2018}. In particular, the control of antiferromagnetic textures offers a pathway toward ultrafast and energy-efficient device operation \cite{Gomonay2016,Shiino2016}. Among these textures, DWs stand out as nanoscale information carriers that combine stability with high mobility, making them promising elements for logic and memory applications \cite{Hedrich2021}. Whereas static interactions such as the DMI define the DW structure and chirality, dynamical mechanisms - most notably magnetic damping - govern their motion, with direct consequences for speed, switching efficiency, stability, and technological applicability.

Within the Landau–Lifshitz framework, magnetic damping is typically characterized by the Gilbert damping constant \cite{Gilbert2004}, whose phenomenological form has been thoroughly explored in FMs, for instance via ferromagnetic resonance linewidth measurements, showing excellent agreement between theory and experiment \cite{Lock1966,Heinrich2002,Tserkovnyak2005,Flovik2015,Haertinger2015,Schoen2016,Hauser2016,Schmidt2020}. Despite these advances, a systematic understanding of how damping governs the dynamics of magnetic textures in antiferromagnets remains elusive. Particularly intriguing is chiral damping - a dissipation mechanism whose magnitude depends on the handedness of magnetic textures such as DWs, skyrmions, and magnons. Recent studies on spin pumping in two-sublattice magnets \cite{Kamra2017,Liu2017,Kamra2018} and on CD in FMs \cite{Jue2016,Ganguly2021,Safeer2022,Akosa2016,Freimuth2017,Akosa2018,Kim2018,Akosa2024} have laid important groundwork. In many respects, CD is the dynamical analogue of the DMI: while DMI imprints chirality on the static magnetic energy landscape, CD governs the dynamical chirality of spin textures. For DWs and skyrmions, DMI stabilizes their structure, whereas CD controls their mobility and response to external drives.
Despite its importance, direct experimental evidence of CD remains elusive. The main challenges are twofold: first, extracting magnetic damping parameters directly from DW dynamics is technically difficult, making unambiguous identification challenging; second, chirality-dependent asymmetries in DW velocity - often interpreted as signatures of CD - can also arise from other sources, such as a chiral gyromagnetic ratio \cite{Freimuth2017,Akosa2018}.

In this work, we investigate the role of CD in AFM DW dynamics, 
highlighting the interplay between tunable sublattice and cross-sublattice contributions and the relativistic Lorentz contraction of the AFM DW width. By controlling chiral asymmetry, for example, via ac gate-voltage modulation of the Rashba SOI, we uncover novel dynamical regimes, including oscillatory motion and chirality switching, which provide experimentally accessible signatures of CD in AFM systems. Furthermore, by exploiting the relativistic constriction of the AFM domain wall at high driving currents, we propose an experimentally viable protocol to obtain both qualitative and quantitative estimates of the CD contribution. Our results demonstrate how this approach can provide direct experimental signatures of CD, paving the way for energy-efficient, programmable spintronic devices with reconfigurable textures.

\textit{Model}.---We consider a one-dimensional uniaxial AFM DW in a system with broken inversion symmetry, composed of two magnetic sublattices $a$ and $b$. The local magnetization is expressed as ${\bf M}^{(i)}({\bm r}, t) = M_{\rm s}{\bf m}^{(i)}({\bm r}, t)$ with $i = a, b$, where $M_{\rm s}$ denotes the saturation magnetization and ${\bf m}^{(i)}({\bm r}, t)$ are unit vectors along the local spin direction. 
Without loss of generality, we include both Rashba and Dresselhaus SOIs, inherent to systems with broken interfacial and bulk inversion symmetries, respectively~\cite{Dresselhaus1955,Bychkov1984}. The dynamics of ${\bf m}^{(i)}$ are then governed by the Landau-Lifshitz-Gilbert-Slonczewski equation.
\begin{eqnarray}\label{eq:cllg0} \nonumber
\partial_t {\bf m}^{(i)} &=& \gamma{\bf H}_{\rm eff}^{(i)}\times {\bf m}^{(i)} + \alpha_{\rm eff}^{(i)}{\bf m}^{(i)}\times\partial_t {\bf m}^{(i)} \\
&-&  b_J^{(i)} \partial_x{\bf m}^{(i)} + \beta b_J^{(i)} {\bf m}^{(i)}\times \partial_x{\bf m}^{(i)} \nonumber\\ 
&+& \gamma H_{\rm sh}^{(i)}{\bf m}^{(i)}\times({\bf y}\times{\bf m}^{(i)}) \nonumber\\
&-& \alpha_{\rm sp}^{i \bar{i}} {\bf m}^{(i)} \times[({\bf m}^{(\bar{i})}\times \partial_t {\bf m}^{(\bar{i})} )\times {\bf m}^{(i)}],
\end{eqnarray}
where, $\gamma$ is the gyromagnetic ratio and ${\bf H}_{\rm eff}^{(i)}$ is the effective field arising from exchange, anisotropy, DMI, and external magnetic fields. The magnitude of the intra-sublattice damping term is given by
\begin{equation}
\alpha_{\rm eff}^{(i)} = \alpha_0 + \alpha_{\rm c}^{(i)}(x,t),
\end{equation}
where $\alpha_0$ denotes the conventional Gilbert damping, and 
$\alpha_{\rm c}^{(i)}(x,t)$ accounts for the nonlocal CD, as formulated in~\cite{Akosa2024}, and is expressed as
\begin{eqnarray}\label{eq:chiral}
\alpha_{\rm c}^{(i)}(x,t)
&=& \Lambda_{\rm R}^{(i)}\mathcal{L}^{(i)}_{zx,x}(x,t)
+\Lambda_{\rm D}^{(i)}\mathcal{L}^{(i)}_{zy,x}(x,t).
\end{eqnarray}
Here, $\Lambda_{{\rm R}({\rm D})}^{(i)}$ are material parameters in units of length that quantify the strength of the Rashba (Dresselhaus) SOI contribution to CD, and $\mathcal{L}^{(i)}_{\mu\nu,\gamma}=m^{(i)}_{\mu}\partial_{\gamma}m^{(i)}_{\nu}-m^{(i)}_{\nu}\partial_{\gamma}m^{(i)}_{\mu}$ is the Lifshitz invariant~\cite{Bogdanov2001,Dzyaloshinsky1964}.
The inter-sublattice damping ($\propto \alpha_{\rm sp}^{i\bar{i}}$, with $\bar{a} (\bar{b})\equiv b (a)$) originates from spin pumping due to the dynamics of the opposite sublattice \cite{Tserkovnyak2002a, Brataas2002, Tserkovnyak2002b, Heinrich2003, Kamra2018, Yuan2019, Pogoryelov2020}.
We include current-induced adiabatic and non-adiabatic spin-transfer torques (STTs) and spin-orbit torque (SOT), corresponding to the third, fourth, and fifth terms on the right-hand side of Eq.~(\ref{eq:cllg0}). The adiabatic STT magnitude is given by $b_{J}^{(i)} = \mu_{\rm B} P^{(i)} j_e /(eM_{\rm s})$, where $P^{(i)}$ is the spin polarization, $j_e$ is the current density, $e$ is the elementary charge, $\mu_{\rm B}$ is Bohr magneton, and $\beta$ is the nonadiabaticity parameter. The damping-like SOT arising from spin Hall effect is characterized by 
$H_{\rm sh}^{(i)} = \hbar \theta_{\rm sh}^{(i)}j_e/(e\mu_0 M_{\rm s}t_{\rm F})$, where $\hbar$ is the reduced Planck's constant, $\theta_{\rm sh}^{(i)}$ is the spin Hall angle for polarization along $\mathbf{y}$, $\mu_0$ is the vacuum permeability, and $t_{\rm F}$ is the film thickness. Following the standard procedure for AFMs \cite{Hals2011,Tveten2013, Tveten2016, Shiino2016, Gomonay2018, Lund2020}, we rewrite Eq.~(\ref{eq:cllg0}) in terms of the averaged magnetization ${\bf m} = \frac{1}{2}({\bf m}^{(a)} + {\bf m}^{(b)})$ and N\'eel vector ${\bf n} = \frac{1}{2}({\bf m}^{(a)} - {\bf m}^{(b)})$. In the exchange limit, where the AFM exchange stiffness $\bar{a}$ dominates all other energy scale, we derive a general equation of motion for $\bf{n}$ by eliminating $\bf{m}$ as detailed in Eq.~(5) of the Supplemental Material (SM). This formulation incorporates all the possible sublattice asymmetries STT, SOT, spin pumping, CD, and their time derivatives.

To elucidate the impact of CD on chiral AFMs, we focus on a one-dimensional AFM DW aligned in the $\bf x$-axis, with the $\bf z$-direction as easy axis. The N\'eel vector is parametrized as ${\bf n} = (\sin\theta\cos\phi, \sin\theta\sin\phi, \cos\theta)$~\cite{Schryer1974}, where $\phi = \phi(t)$ is the DW tilt angle and $\theta(x,t) = 2\tan^{-1}\{\exp[(x - X(t))/\lambda]\}$. Here, $X(t)$ and $\lambda=a_0\sqrt{A/K}$ denote the DW center and width, respectively ($a_0$ is lattice constant, $A$ is the effective FM exchange constant, and $K$ is easy-axis anisotropy energy). 
For simplicity, we neglect distortions of $\lambda$ and nonlinear driving effects, so that $\dot{\bf n} = \dot{X} \partial_X{\bf n} + \dot{\phi}\partial_\phi{\bf n}$ and $\ddot{\bf n} = \ddot{X} \partial_X{\bf n} + \ddot{\phi}\partial_\phi{\bf n}$ \cite{Tveten2013}. Projecting the equation of motion for $\bf n$ onto $\partial_x{\bf n}$ and ${\bf n} \times \partial_x{\bf n}$, and integrating over $x$, we obtain the coupled equations for $X$ and $\phi$ shown in Eq.~(11) in SM.

\textit{Results}.---As a demonstration of the impact of CD, we focus on a case with finite RSOI and the associated interfacial DMI of strength $D_{\rm I}$, relevant for materials such as synthetic AFMs~\cite{Yang2015,Yang2017, Yang2019}. Using the DW ansatz for $\bf n$, the sublattice-equivalent DMI field is taken as $\mathcal{H}_{\rm DM}=D_{\rm I}\cos\phi/(\mu_0 M_{\rm s}\cosh[(x-X)/\lambda])$, while the nonlocal CD reads $\alpha_{\rm c}^{(i)}=\Lambda^{(i)}_{ \rm R}\cos\phi/(\lambda\cosh[(x-X)/\lambda])$ with $\Lambda^{(i)}_{\rm R}$ being the strength of CD proportional to that of RSOI. 
Defining $\Lambda_{\rm R}^{\bf m,n}=\frac{1}{2}(\Lambda_{\rm R}^{(a)}\pm\Lambda_{\rm R}^{(b)})$, we take $\text{sign}(D_{\rm I})=-\text{sign}(\Lambda_{\rm R}^{\bf m})$, since a decrease in DMI energy should correspond to an increase in the damping~\cite{Akosa2024}, wheras the sign of $\Lambda^{\bf n}_{\rm R}$ is determined by sublattice material properties and enviroment. 
We consider an applied dc-current to induce sublattice-equivalent STT and SOT with $b_J^{(i)}=b_J, H_{\rm sh}^{(i)}=H_{\rm sh}$, while the CD constants $\Lambda^{\bf m,n}_{\rm R}(t)$ oscillate in time at the frequency of an applied ac gate voltage. For simplicity, spin pumping is neglected ($\alpha_{\rm sp}^{i\bar{i}}=0$). Under these assumptions, the coupled equations for $q_{\mu=X,\phi}=X(t), \phi(t)$ take the form
\begin{equation}\label{eq:nsl}
\mathcal{M}_{\mu\nu} \ddot{q}_\nu + \mathcal{N}_{\mu\nu} \dot{q}_\nu  = \mathcal{F}_{\mu},
\end{equation}
with summation over repeated indices implied. Here the mass ($\mathcal{M}_{\mu\nu}$), drag ($\mathcal{N}_{\mu\nu}$) and force ($\mathcal{F}_{\mu}$) tensors are defined as
\begin{widetext}
\begin{eqnarray}
\mathcal{M}_{XX} &=& -\frac{1}{\lambda}\mathcal{M}_{\phi \phi}= \frac{2}{\gamma\lambda}\Big[1 + \alpha_0^2+2\bar{\alpha}^{\mathbf{m}}_{\rm c} \cos\phi\Big(\alpha_0 + \frac{16}{3\pi^2}\bar{\alpha}^{\mathbf{m}}_{\rm c}\cos\phi\Big)\Big],\ 
\mathcal{M}_{X\phi}=\lambda\mathcal{M}_{\phi X}=\frac{-2\bar{\alpha}^{\mathbf{n}}_{\rm c}}{\gamma}\cos\phi,\nonumber\\
\mathcal{N}_{XX}&=&\frac{2\bar{a}}{\lambda}\Big(\alpha_0 + \bar{\alpha}^{\mathbf{m}}_{\rm c} \cos\phi\Big)+\frac{4}{\gamma\lambda}\Big(\alpha_0 + \frac{32}{3\pi^2}\bar{\alpha}^{\mathbf{m}}_{\rm c}\cos\phi \Big)\dot{\bar{\alpha}}^{\mathbf{m}}_{\rm c} \cos\phi,\nonumber\\
\mathcal{N}_{\phi \phi}&=&-\lambda \mathcal{N}_{XX}+\pi H_{\rm D}\cos\phi\Big(\alpha_0 +\frac{32}{3\pi^2}\bar{\alpha}^{\mathbf{m}}_{\rm c}  \cos\phi\Big)-\frac{32}{3\pi}\bar{\alpha}^{\mathbf{m}}_{\rm c} H_{\rm D}\sin^2\phi,\nonumber\\
\mathcal{N}_{X\phi}&=&-\frac{2}{\gamma}\dot{\bar{\alpha}}^{\mathbf{n}}_{\rm c} \cos\phi+\frac{\pi}{2}H_{\rm sh}\sin\phi\Big(\alpha_0+\frac{32}{3\pi^2}\bar{\alpha}^{\mathbf{m}}_{\rm c}\cos\phi\Big)
+2\bar{\alpha}^{\mathbf{m}}_{\rm c}\sin\phi\Big(\frac{\beta b_J }{\lambda\gamma}+\frac{4}{\pi}H_{\rm sh} \cos\phi\Big),\nonumber\\
\mathcal{N}_{\phi X}&=&-\frac{2}{\gamma\lambda}\dot{\bar{\alpha}}^{\mathbf{n}}_{\rm c} \cos\phi+\frac{\pi}{2\lambda}H_{\rm sh}\sin\phi\Big(\alpha_0+\frac{16}{3\pi^2}\bar{\alpha}^{\mathbf{m}}_{\rm c}\cos\phi\Big),\nonumber\\
\mathcal{F}_{X}&=& \bar{a}\gamma\Big(\frac{2\beta b_J}{\lambda\gamma}+\pi H_{\rm sh} \cos\phi\Big)+\dot{\bar{\alpha}}^{\mathbf{m}}_{\rm c}\cos\phi\Big(\frac{2\beta b_J }{\lambda\gamma}+\frac{8}{\pi}H_{\rm sh}\cos\phi\Big),\
\mathcal{F}_{\phi}=-\frac{16}{3\pi}\dot{\bar{\alpha}}^{\mathbf{m}}_{\rm c} H_{\rm D}\sin 2\phi-\bar{a}\gamma\pi H_{\rm D}\sin\phi.\label{tensor}
\end{eqnarray}
\end{widetext}
We define the dimensionless CD constants $\bar{\alpha}^{\mathbf{m,n}}_{\rm c}\equiv\pi\Lambda^{\mathbf{m,n}}_{\rm R}/(4\lambda)$  and the DMI-induced field $H_{\rm D} = D_{\rm I}/(\mu_0 M_s\lambda)$.
Equation~(\ref{eq:nsl}) can be interpreted as describing the classical motion of a massive particle subject to external and drag forces. Our results indicate that the coupling between $X$ and $\phi$ of an AFM DW is primarily mediated by CD, together with STT and SOT.  This coupling gives rise to rich dynamical phenomena, essential for achieving precise control in AFM spintronics \cite{Tveten2013, Gomonay2014, Tveten2014, Shiino2016}, providing a pathway through which the dynamics of one degree of freedom can influence the other and break the inherent symmetry of their otherwise independent dynamics. Importantly, in Eq.~(\ref{tensor}), the effective mass tensor $\mathcal{M}_{\mu\nu}$ is renormalized by CD ($\propto\bar{\alpha}^{\bf m,n}_{\rm c}$), which modifies how energy is stored within the texture and impacts the inertial properties of an AFM DW. By contrast, the coupling terms induced by STT ($\propto b_J$) and SOT ($\propto H_{\rm sh}$) are proportional to $\sin\phi$ and thus vanish for a pure N\'eel DW with $\phi(t=0) = 0,\pi$. 
In particular, in the absence of CD, the dynamics of $X$ and $\phi$ of a N\'eel-type AFM DW are fully decoupled ($\mathcal{M}_{\mu \nu} = \mathcal{N}_{\mu \nu} = 0$ for $\mu\ne \nu$), reducing to the standard AFM DW dynamics \cite{Hals2011, Tveten2013, Tveten2016, Shiino2016, Gomonay2018, Lund2020}. Hence, CD, introduces a mechanism for more intricate and diverse dynamical behaviors in AFMs.

\textit{Perturbative solution}.---Analytical solutions for the DW velocity $v(t)=\dot{X}(t)$ and tilt angle $\phi(t)$ can be obtained via a perturbative expansion in the CD. We assume the CD parameters oscillate as $\bar{\alpha}^{\bf m,n}_{\rm c}=\bar{\alpha}^{\bf m,n}_{\rm c,0}\sin(\Omega t)$ and $\dot{\bar{\alpha}}^{\bf m,n}_{\rm c}=\Omega\bar{\alpha}^{\bf m,n}_{\rm c,0}\cos(\Omega t)$, where $\Omega$ is the frequency of the applied gate voltage. From Eq.~(\ref{eq:nsl})--(\ref{tensor}), both $v(t)$ and $\phi(t)$ 
are expected to exhibit oscillatory behaviors driven by CD. To proceed, we substitute the resonant ansatz
\begin{eqnarray}
v(t)&\approx&v_0+v_s\sin(\Omega t)+v_c\cos(\Omega t),\nonumber\\
\phi(t)&\approx&\phi_0+\phi_s\sin(\Omega t)+\phi_c\cos(\Omega t),\label{ansatz}
\end{eqnarray}
into Eq.~(\ref{tensor}). However, this generates additional terms proportional to $\sin^2(\Omega t)$, $\cos^2(\Omega t)$, and $\sin(2\Omega t)$, indicating that the exact solution for $v(t)$ and $\phi(t)$ generally involves a combination of $\sin(\Omega t)$, $\cos(\Omega t)$, and their higher harmonics. Since the typical values of $\bar{\alpha}^{\bf m,n}_{\rm c,0}$ are small \cite{Akosa2018}, the equations of motion can be solved perturbatively in $\bar{\alpha}^{\bf m,n}_{\rm c,0}$. The coefficients $v_s, v_c, \phi_s, \phi_c$ in Eq.~(\ref{ansatz}) are linear in CD and thus directly proportional to $\bar{\alpha}^{\bf m,n}_{\rm c,0}$, while the neglected terms involve higher-order combinations of $\sin(\Omega t)$ and $\cos(\Omega t)$, and are therefore higher-order in CD. 
Keeping only terms up to first order in CD, Eq.~(\ref{eq:nsl}) reduces to two equations of the form $0\approx c_1+c_2\cos(\Omega t)+c_3\sin(\Omega t)$ and $0\approx c_4+c_5\cos(\Omega t)+c_6\sin(\Omega t)$, where $c_j$ are functions of the six unknown variables in Eq.~(\ref{ansatz}). Solving $c_j=0$ with $j=1, ...,6$ yields the pertubative solutions for these variables (see SM Sec.~II). As a result, the time-independent components of $v$ and $\phi$ are
\begin{eqnarray}
v_0&=&\frac{\beta b_J}{\alpha_0}+\frac{\pi\gamma H_{\rm sh}\lambda}{2\alpha_0},\ \phi_0=0,\label{v}
\end{eqnarray}
while the amplitudes of the oscillating components are
\begin{eqnarray}
&&\delta v\equiv\sqrt{v_s^2+v_c^2}\approx\frac{\bar{a}\alpha^{\mathbf{m}}_{\rm c,0}\gamma v_0 }{\Omega},\label{amp}\\
&&\delta\phi\equiv\sqrt{\phi_s^2+\phi_c^2}
\approx\frac{4\alpha_{\rm c,0}^{\mathbf{n}}\Omega v_0}
{\pi\alpha_0\gamma H_{\rm sh} v_0
+2\lambda(\pi\gamma^2\bar{a}H_{\rm D}+2\Omega^2)}.\nonumber
\end{eqnarray}
Here $\delta v$ and $\delta\phi$ are expanded up to the lowest order of Gilbert damping $\alpha_0$. The phase shifts relative to the gate voltage, $\tan^{-1}(v_c/v_s)$ and $\tan^{-1}(\phi_c/\phi_s)$, can also be determined but are not shown here. Equation~(\ref{v}) shows that the time-averaged DW velocity $v_0$ consists of a spin drift velocity contribution coming from nonadiabatic STT and a part induced by SOT via the AFM exchange torque~\cite{Yang2015,Yang2017, Blasing2018,Cohen2020,Yang2019}. 
Importantly, Eq.~(\ref{amp}) explicitly demonstrates the dependencies of the oscillation amplitudes of $v$ and $\phi$ on the CD magnitudes $\alpha^{\bf m,n}_{\rm c,0}$, frequency of applied gate voltage $\Omega$, external current via STT and SOT, the AFM exchange stiffness $\bar{a}$, and the DMI field $H_{\rm D}$. Notably, the symmetric and asymmetric components of sublattice CDs, $\alpha^{\bf m}_{\rm c,0}$ and $\alpha^{\bf n}_{\rm c,0}$, respectively induce oscillations in DW velocity and tilt angle. These results provide a direct route for experimental detection and quantification of CD in synthetic AFMs.

\begin{figure}[h]
\centering
\includegraphics[scale=0.47]{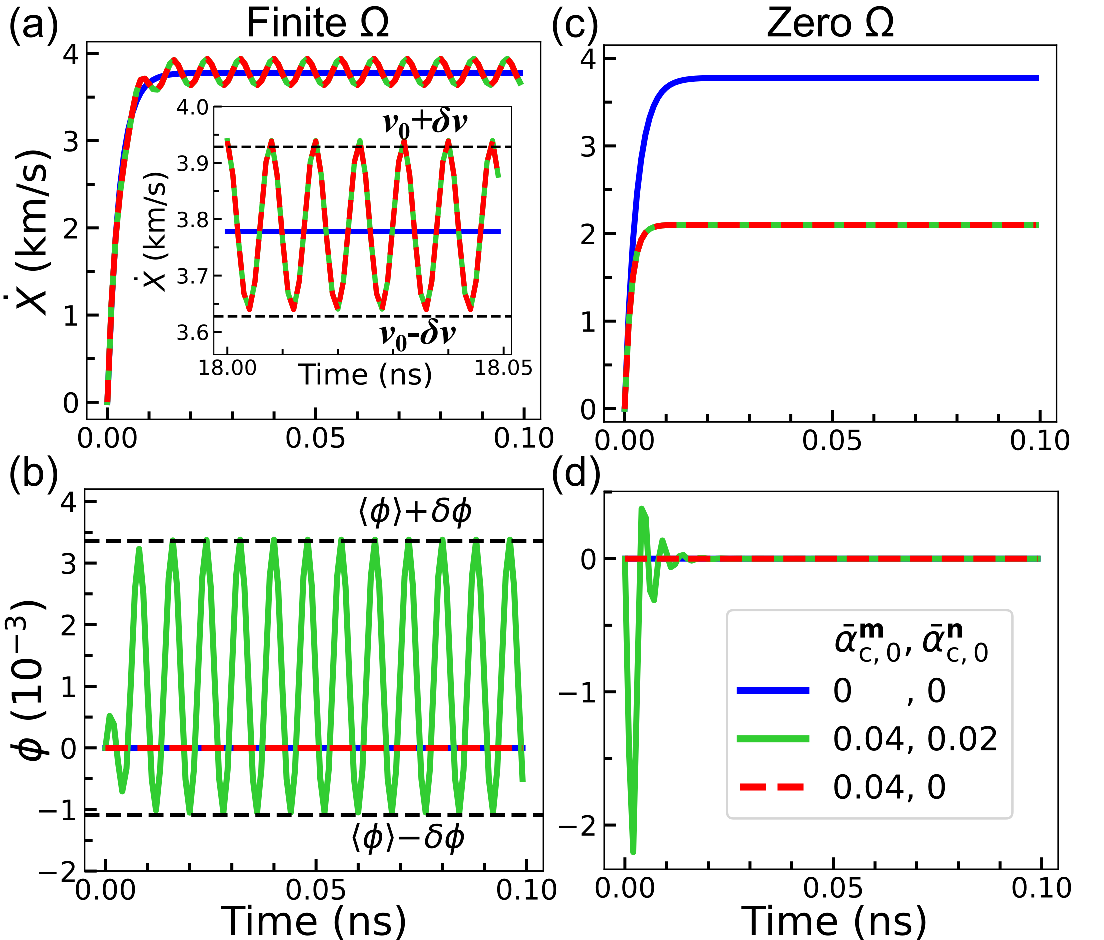}
\caption{Numerical results of DW velocity and tilt angle under both SOT and STT for CDs with (a)--(b) a finite frequency $\Omega$, and (c)--(d) zero frequency. The inset of (a) shows an enlarged view. The horizontal black dotted lines in the inset of (a) denote $v_0\pm \delta v$, while those in (b) denote $\langle\phi\rangle\pm\delta\phi$.}
\label{Fig2}
\end{figure}
\textit{Numerical result.}---To validate the perturbative solutions, Eq.~(\ref{eq:nsl})--(\ref{tensor}) are numerically integrated using the fourth-order Runge-Kutta (RK) algorithm with parameters $\alpha_0=0.05$, $\beta=0.2$, $\bar{\alpha}^{\mathbf{m}}_{\rm c,0}=0.04$, $\bar{\alpha}^{\mathbf{n}}_{\rm c,0}=0.02$, $\theta^{(i)}_{\rm sh}=0.15$, $P^{(i)}=0.6$, $j_e=2\times 10^{12}$~A/m$^2$, $H_{\rm D}=-1$~T, $\bar{a}=40$~T, $\Omega=2\pi\times 1.125$ THz, $\mu_0M_{\rm s}=1$~T, $a_0=0.5$~nm, $\lambda=10 a_0$, and $t_{\rm F}=2$~nm. 
From Eq.~(\ref{v}), the time-averaged DW velocity part induced by STT is $\beta b_J/\alpha_0\approx 0.35$~km/s, whereas the part induced by SOT is $\pi\gamma H_{\rm sh}\lambda/(2\alpha_0)\approx 3.4$~km/s, showing that SOT is much more efficient in driving the DW motion.

When both SOT and STT are present, Fig.~\ref{Fig2}~(a)--(b) show the RK results of DW velocity and tilt angle with oscillating CDs. Although only data up to 0.1 ns are displayed, calculations were performed up to 20 ns, confirming that the dynamics has stabilized after approximately 0.1 ns. 
The blue, green, and red curves respectively correspond to (i) zero $\bar{\alpha}^{\bf m,n}_{\rm c,0}$, (ii) finite $\bar{\alpha}^{\bf m,n}_{\rm c,0}$, and (iii) finite $\bar{\alpha}^{\bf m}_{\rm c,0}$ but zero $\bar{\alpha}^{\bf n}_{\rm c,0}$ (i.e., without sublattice CD asymmetry). 
The oscillatory behavior in DW velocity and/or tilt angle emerges only when CD is present, with a period close to $\Omega$.
In the inset of Fig.~\ref{Fig2}~(a), the analytical value $v_0\approx 3.78$~km/s from Eq.~(\ref{v}) closely matches the numerical average from 10 to 20 ns, while the black horizontal dashed lines representing $v_0\pm\delta v$ capture the oscillation range of RK results very accurately. 
In Fig.~\ref{Fig2}~(b), there is a finite averaged tilt angle $\langle\phi\rangle\approx 1.1\times10^{-3}$ which can not be captured by our first-order perturbation theory which predicts $\phi_0=0$; however, when taking $\langle\phi\rangle$ extracted from the RK result as the origin, the range $\langle\phi\rangle\pm\delta\phi$ with $\delta\phi$ calculated by Eq.~(\ref{amp}) again matches the numerical oscillation well. 
In general, the oscillating RSOI induced by an ac gate voltage will result in both the oscillations of DMI and CD. We have calculated the dynamics of $X$ and $\phi$ in the presence of both oscillating DMI and CD as shown in SM Sec.~III. The DW dynamics is similar to Figs.~\ref{Fig2}~(a)--(b).
These results confirm our prediction in the previous section: a finite $\alpha^{\bf m}_{\rm c,0}$ (red and green curves) and $\alpha^{\bf n}_{\rm c,0}$ (green curve) induce oscillations of DW velocity and tilt angle, respectively, providing a clear experimental signature of CD. While the oscillation amplitude of $\phi$ is small with our model parameters, Eq.~(\ref{v})--(\ref{amp}) show that by tuning $b_J, H_{\rm sh}, \lambda, \bar{a}, H_{\rm D}$, and $\Omega$, it may allow for experimentally  observable tilt-angle oscillations.

To compare the DW dynamics induced by oscillating and static CDs, Fig.~\ref{Fig2}~(c)--(d) show the RK results with $\Omega=0$. Here Eq.~(\ref{amp}) is inapplicable due to divergence; instead, starting from Eq.~(\ref{tensor}) with $\dot{\bar{\alpha}}^{\bf m,n}_{\rm c}=0$ and $\bar{\alpha}^{\bf m,n}_{\rm c}=\bar{\alpha}^{\bf m,n}_{\rm c,0}$, a steady solution exists with $\ddot{X}(t)=0$ and $\phi(t)=0$, yielding $v=v_0\alpha_0/(\alpha_0+\bar{\alpha}^{\bf m}_{\rm c,0})$ with $v_0$ defined in Eq.~(\ref{v}). 
This shows that the static CD renormalizes the effective Gilbert damping by $\bar{\alpha}^{\bf m}_{\rm c,0}$. Indeed, the terminal DW velocities in green and red curves of Fig.~\ref{Fig2}~(c) match this value, and Fig.~\ref{Fig2}~(d) shows $\phi=0$ for all cases after a transient regime. Thus a static gate voltage modifies the damping but does not induce oscillatory dynamics.

\begin{figure}[h]
\centering
\includegraphics[scale=0.46]{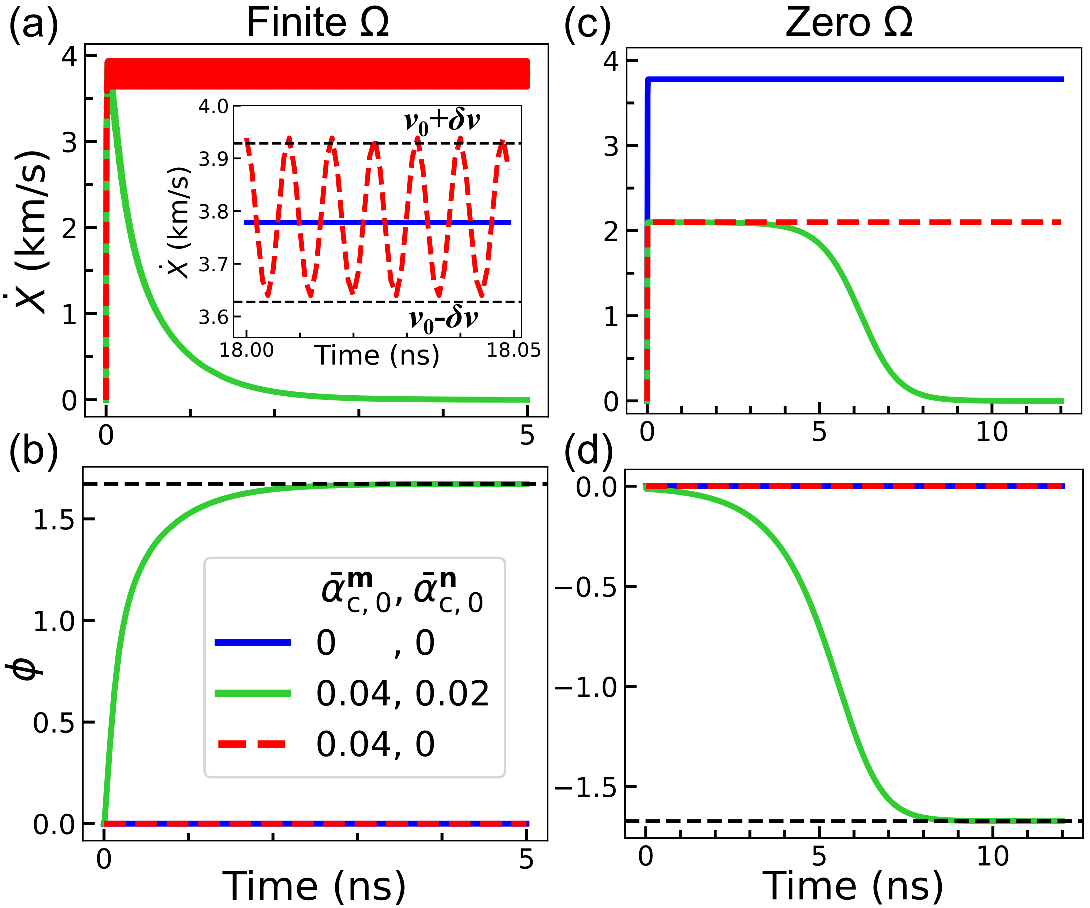}
\caption{Numerical results of DW velocity and tilt angle without DMI, $H_{\rm D}=0$. The inset in (a) shows an enlarged view.}
\label{Fig3}
\end{figure}
We now consider the case with vanishing DMI ($H_D=0$) but with finite CDs. Figure~\ref{Fig3} shows the RK results under these conditions.  For $\Omega \ne 0$ [Figs.~\ref{Fig3}~(a)--(b)], the key difference compared to Figs.~\ref{Fig2}~(a)--(b) is that when both $\bar{\alpha}^{\bf m,n}_{\rm c}$ are finite (green curve), the terminal DW velocity will drop to zero, and the tilt angle will be rotated from zero to a value close to $\pi/2$. 
This behavior can be understood by setting $H_{\rm D}=0$ in Eq.~(\ref{tensor}). A steady solution exists with $\dot{X}(t)\approx 0$ and $\dot{\phi}(t)\approx 0$ when $\mathcal{F}_X\approx 0$ , which can be approximately satisfied when $\phi=\pm \cos^{-1}[-2\beta b_J/(\pi\gamma H_{\rm sh}\lambda)]$. The black dotted line in Fig.~\ref{Fig3}~(b) shows the positive branch of this solution, in good agreement with the RK result. Notably, when non-adiabatic STT is small relative to SOT, 
$\phi\approx \pm\pi/2$, indicating that the initial N\'eel DW switches to a Bloch DW due to finite 
$\bar{\alpha}^{\bf m,n}_{\rm c}$ in the absence of DMI.

On the other hand, for $\Omega=0$  [Figs.~\ref{Fig3}~(c)--(d)], the behavior is similar, but the terminal $\phi$ for green curve takes the negative value. The sign is determined by the initial angular accelerations in the two cases. Furthermore, reversing the sign of $\bar{\alpha}^{\bf n}_{\rm c,0}$ flips the terminal $\phi$ for the green curves in both (b) and (d). 
When the sublattice asymmetry of CD is absent ($\bar{\alpha}^{\bf n}_{\rm c}=0$), the red curves in Figs.~\ref{Fig2}--\ref{Fig3} show that $\phi$ will stay zero regardless of the presence of DMI. In this case, $\mathcal{M}_{X\phi}=\mathcal{M}_{\phi X}=0$ in Eq.~(\ref{tensor}), and the initial conditions $\dot{X}(0)=\dot{\phi}(0)=\phi(0)=0$ yield $\mathcal{F}_{\phi}(0)=0$, leading to $\ddot{\phi}(0)=0$. Consequently $\phi(t) =0$ at all the times. Therefore, the switching of DW tilt angle by CD in the absence of DMI as shown in Fig.~\ref{Fig3} requires finite asymmetric CDs in magnetic sublattices.

\textit{Discussion}.---Our work highlights the central role of CD in shaping AFM DW dynamics. 
Sublattice-asymmetric CD generates off-diagonal components of the DW mass tensor, 
thereby coupling translational and rotational degrees of freedom and producing nontrivial dynamics. 
Gate-controlled modulation of the RSOI enables tunable oscillations of the DW velocity and tilt angle, 
while, remarkably, asymmetric CD alone permits reversible chirality switching between N\'eel- and 
Bloch-type DWs even in the absence of DMI. These effects provide clear and experimentally 
accessible signatures of CD and can be probed using advanced magnetic imaging techniques 
such as x-ray magnetic circular dichroism or spin-polarized scanning tunneling microscopy. 
Our analytical theory, supported by numerical simulations, demonstrates the robustness of these 
phenomena and identifies CD as a versatile and tunable control parameter for AFM DW dynamics. 
In this sense, CD plays a role analogous to a dynamical DMI, enabling time-dependent 
and reversible control of magnetic chirality beyond what is achievable through static interactions alone. 
More broadly, these results extend the understanding of dynamical chirality in magnetic textures and 
point toward new routes for reconfigurable spintronic functionalities. 

Finally, we propose a practical experimental strategy to quantitatively extract the CD contribution 
in antiferromagnets using resonance-linewidth measurements~\cite{Mekonnen2011,Mizukami2011}. 
Unlike the conventional Gilbert damping parameter $\alpha_0$, 
which is independent of the domain-wall width $\lambda$, the chiral contribution $\alpha_c$ scales 
with the wall gradient, $\alpha_c \propto 1/\lambda$. By measuring the antiferromagnetic resonance 
linewidth in low-driving regimes, where the DW remains wide, and in high-driving regimes, 
where relativistic Lorentz contraction compresses the wall, one selectively modulates the chiral 
dissipation channel while leaving $\alpha_0$ unchanged. The resulting difference in effective 
damping directly yields a quantitative measure of the chiral contribution.

\textit{Conclusion}.---We study the effect of CD on the DW dynamics in AFMs. We derive the general coupled equations of DW velocity and tilt angle under the influences of sublattice STT, SOT, spin pumping effect, and CDs. Focusing on the case of Rashba SOI, we calculate the time-dependent DW velocity and tilt angle both analytically and numerically under oscillating CDs induced by an ac gate voltage. We find the oscillating wall velocity and tilt angle induced by finite sublattice averaged and asymmetric CDs, respectively. Our work provides a prediction for experiments to directly measure the existence and magnitude of CD in AFMs.

\textit{Acknowledgement.}— C.A.A. and M-K.L. contributed equally to this work.

\clearpage
\onecolumngrid

\title{Supplemental Material: Chiral Damping-Induced Chirality Switching and Control of Domain Walls in Antiferromagnets}

{\centering{\large\textbf{Supplemental Material: Chiral Damping-Induced Chirality Switching and Control of Domain Walls in Antiferromagnets}}}\\ \\
Collins Ashu Akosa$^{1,2}$, Mu-Kun Lee$^1$, Aur\'elien Manchon$^3$, and Masahito Mochizuki$^1$\\
{\footnotesize \textit{$^1$Department of Applied Physics, Waseda University, Okubo, Shinjuku-ku, Tokyo 169-8555, Japan\\
		$^2$Department of Theoretical and Applied Physics, African University of Science and Technology (AUST), Km 10 Airport Road, Galadimawa, Abuja F.C.T, Nigeria\\
		$^3$Aix-Marseille Universit\'e, CNRS, CINaM, Marseille, France}}

\tableofcontents
\section{Equation of motion for the N\'eel vector}
By following the standard procedure for AFMs \cite{Hals2011,Tveten2013, Tveten2016, Shiino2016, Gomonay2018, Lund2020}, we rewrite Eq.~(1) in the main text in terms of the magnetization vector ${\bf m} = \frac{1}{2}({\bf m}^{(a)} + {\bf m}^{(b)})$ and N\'eel vector ${\bf n} = \frac{1}{2}({\bf m}^{(a)} - {\bf m}^{(b)})$. In our linear response analysis, we preserve terms up to linear order in the driving force to obtain 
\begin{eqnarray}
\dot{\bf m} &=&  \Big(\gamma{\bf f}_{\bf n} - (\alpha^{\bf m}_{\rm eff} - \alpha^{\bf m}_{\rm  sp}) \dot{\bf n} 
- \beta b^{\bf m}_J{\bf n}' - \gamma{H}^{\bf m}_{\rm sh}{\bf y}\times{\bf n} - b^{\bf n}_J {\bf n}\times{\bf n}' \Big) \times{\bf n} + \boldsymbol{\tau}_{\rm nl}^{\bf m},\label{eq:mym}\\
\dot{\bf n} &=& \Big(\gamma {\bf f}_{\bf m} - (\alpha^{{\bf m}}_{\rm eff}+\alpha^{{\bf m}}_{\rm  sp})\dot{{\bf m}}-b^{{\bf m}}_J{\bf n}\times{\bf n}'-(\alpha^{\bf n}_{\rm eff}-\alpha^{\bf n}_{\rm sp})\dot{\bf n} -\beta b^{\bf n}_J{\bf n}'-\gamma H^{\bf n}_{\rm sh}{\bf y}\times{\bf n} \Big)\times{\bf n} + \boldsymbol{\tau}_{\rm nl}^{\bf n}, \label{eq:myn}
\end{eqnarray}
where $\boldsymbol{\tau}_{nl}^{\bf m}$ and $\boldsymbol{\tau}_{nl}^{\bf n}$ represent higher-order terms  neglected for this work. We have defined the following notations: $\dot{\bf n} \equiv \partial_t{\bf n}$ and ${\bf n}' \equiv \partial_x{\bf n}$,
$\alpha^{{\bf m}({\bf n})}_{\rm eff} \equiv \frac{1}{2}(\alpha_{\rm eff}^{(a)} \pm \alpha_{\rm eff}^{(b)})$, 
$\alpha_{\rm  sp}^{{\bf m}({\bf n})} \equiv \frac{1}{2}(\alpha_{\rm sp}^{ab} \pm \alpha_{\rm sp}^{ba})$,
${\bf f}_{{\bf m}({\bf n})} \equiv \frac{1}{2}({\bf H}_{\rm eff}^{(a)} \pm {\bf H}_{\rm eff}^{(b)})$, 
$b_J^{{\bf m}({\bf n})} \equiv \frac{1}{2}( b_J^{(a)} \pm b_J^{(b)})$, and
${H}_{\rm sh}^{{\bf m}({\bf n})} \equiv \frac{1}{2}( H_{\rm sh}^{(a)} \pm H_{\rm sh}^{(b)})$. It is important to note that although our analysis assumes equivalent sublattices with similar material parameters, we do not exclude the potential for asymmetry caused by inhomogeneity or chirality, which may be present in real materials. As a result, effective parameters such as $b_J^{\bf n}$, ${H}_{\rm sh}^{\bf n}$, and $\alpha^{\bf n}_{\rm eff}$ may have nonzero values. 
Indeed, it is anticipated that chiral AF DWs will not only exhibit distinct current-induced torques on different sublattices due to their unique environment, which are the origin of the chiral exchange drag for DWs in synthetic AFMs \cite{Gramila1991,Sivan1992}, but also display effects such as chiral spin-pumping and damping within the two sublattices \cite{Kamra2017, Liu2017, Kamra2018}. These combined effects would led to a non-zero value for the effective damping, $\alpha^{\bf n}_{\rm eff}$. 
We derive the expression for {\bf m} by substituting of Eq.~(\ref{eq:mym}) into Eq.~(\ref{eq:myn}) to get
\begin{eqnarray}\label{eq:mag}
{\bf m}  &=& \frac{1}{ \bar{a}} \Big( {\bf n}\times{\bf H}_{{\bf m}} - (\alpha^{{\bf m}}_{\rm eff} +\alpha^{{\bf m}}_{\rm  sp}){\bf f}_{\bf n}  -
\frac{1}{\gamma } (\alpha^{\bf n}_{\rm eff}-\alpha^{\bf n}_{\rm  sp}){\bf n}\times\dot{\bf n} +  \frac{\dot{\bf n} }{\tilde{\gamma}(x,t)}  - H^{\bf n}_{\rm sh}{\bf y} \Big) \times{\bf n}\nonumber  \\ 
&+& \frac{1}{ \bar{a}}\Big(  \frac{b^{{\bf m}}_J }{\gamma} \big( 1+ (\alpha^{{\bf m}}_{\rm eff} +\alpha^{{\bf m}}_{\rm  sp}) \beta \big){\bf n}'  - \frac{b^{\bf n}_J}{\gamma }\big(\beta -  (\alpha^{{\bf m}}_{\rm eff} +\alpha^{{\bf m}}_{\rm  sp}) \big) {\bf n}\times{\bf n}' + (\alpha^{{\bf m}}_{\rm eff} +\alpha^{{\bf m}}_{\rm  sp}) H^{{\bf m}}_{\rm sh} {\bf y}\times{\bf n} \Big) \times{\bf n},
\end{eqnarray}
where 
\begin{equation}\label{eq:gyro}
\tilde{\gamma}(x,t) \equiv  \gamma/\big( 1  +   (\alpha^{\bf m}_{\rm eff} +  \alpha_{\rm  sp}^{\bf m})  (\alpha^{\bf m}_{\rm eff} -  \alpha_{\rm  sp}^{\bf m}) \big).
\end{equation}
In the derivation, we have used ${\bf f}_{\bf m}\approx {\bf H}_{\bf m}-\bar{a}\bf{m}$ where $\bf{H}_{\bf m}$ is the averaged applied magnetic field and $\bar{a}=4J_{\rm AF}/(M_{\rm s}a^3_0)$ with $J_{\rm AF}$ being the AF exchange stiffness in units of energy. As anticipated in systems with broken inversion symmetry and strong SOI, 
the renormalized gyromagnetic ratio in our model varies both in time and across space, i.e., $\tilde{\gamma}= \tilde{\gamma}(x,t)$~\cite{Freimuth2017, Kim2018, Akosa2018}. From the expression of ${\bf m}$ in Eq.~(\ref{eq:mag}), the following key inferences can be made: 
(i) ${\bf m}$ is a variable entirely determined by the spatial and temporal dynamics of Néel vector ${\bf n}$.
(ii) We have the equality ${\bf m}\cdot{\bf n}=0$ as required by the unit length of vector ${\bf m}^{(i)}$, and ${\bf m}$ vanishes at equilibrium when no driving forces are present, aligning with the assumption of equivalent sublattices.
(iii) External currents can induce a small ${\bf m}$ via STT and/or SOT even in AFMs with equivalent sublattices. The induced magnetic moment ${\bf m}$ has been demonstrated to be crucial, particularly in the SOT-induced dynamics of AF DWs. In such cases, it is known to render the exchange torque highly effective in driving the DWs in synthetic AFMs at high speeds up to a few km/s~\cite{Yang2015,Yang2017, Blasing2018,Cohen2020,Yang2019}. 
In particular, the SOT results in a nonzero in-plane magnetization as described by the last term in the second line of Eq.~(\ref{eq:mag}). Consequently, it can be determined that when chiral asymmetries exist between the two sublattices, the third and fifth terms in the first line of Eq.~(\ref{eq:mag}) provide an additional pathway to enhance the resultant velocity for SOT-driven AF DW dynamics. 
In particular, the chiral asymmetry, represented by the third term in the first line of the right-hand side of Eq.~(\ref{eq:mag}), enhances the net magnetization through additional spin canting, thereby renormalizing the DW mass.

The equation of motion for the N\'eel vector {\bf n} is derived (disregarding terms that are quadratic in the driving force, i.e. $\propto \dot{\bf n}^2$ \cite{Tveten2013}) by substituting Eq.~(\ref{eq:mag}) into Eq. (\ref{eq:mym}) to obtain 
\begin{eqnarray}\label{eq:eom_n}
&&\bar{a}\beta b^{\mathbf{m}}_J\mathbf{n}'-\bar{a}\gamma H^{\mathbf{m}}_{\rm sh}\mathbf{n}\times\mathbf{y}-\bar{a}\gamma\mathbf{f}_{\mathbf{n}}+\mathbf{n}\times\dot{\mathbf{H}}_{\mathbf{m}}-(\alpha^{\mathbf{m}}_{\rm eff}+\alpha^{\mathbf{m}}_{\rm sp})\dot{H}^{\mathbf{m}}_{\rm sh}\mathbf{n}\times\mathbf{y}-\frac{1}{\gamma}(\dot{\alpha}^{\mathbf{n}}_{\rm eff}-\dot{\alpha}^{\mathbf{n}}_{\rm sp})\mathbf{n}\times\dot{\mathbf{n}}\nonumber\\
&&+\frac{1}{\gamma}[1+\beta(\alpha^{\mathbf{m}}_{\rm eff}+\alpha^{\mathbf{m}}_{\rm sp})]\dot{b}^\mathbf{m}_J\mathbf{n}'-\frac{1}{\gamma}[\beta-(\alpha^{\mathbf{m}}_{\rm eff}+\alpha^{\mathbf{m}}_{\rm sp})]\dot{b}^{\mathbf{n}}_J\mathbf{n}\times\mathbf{n}'-\dot{H}^{\mathbf{n}}_{\rm sh}\mathbf{y}+\dot{\mathbf{n}}\times\mathbf{H}_{\mathbf{m}}-(\mathbf{n}\cdot\dot{\mathbf{n}}\times\mathbf{H}_{\mathbf{m}})\mathbf{n}\nonumber\\
&&+(\dot{\alpha}^{\mathbf{m}}_{\rm eff}+\dot{\alpha}^{\mathbf{m}}_{\rm sp})(\mathbf{n}\cdot\mathbf{f}_{\mathbf{n}})\mathbf{n}
+(\alpha^{\mathbf{m}}_{\rm eff}+\alpha^{\mathbf{m}}_{\rm sp})(\mathbf{n}\cdot\dot{\mathbf{f}}_{\mathbf{n}})\mathbf{n}+(\alpha^{\mathbf{m}}_{\rm eff}+\alpha^{\mathbf{m}}_{\rm sp})H^{\mathbf{m}}_{\rm sh}{\mathbf{y}}\times\dot{\mathbf{n}}-(\alpha^{\mathbf{m}}_{\rm eff}+\alpha^{\mathbf{m}}_{\rm sp})H^{\mathbf{m}}_{\rm sh}(\mathbf{n}\cdot\mathbf{y}\times\dot{\mathbf{n}})\mathbf{n}\nonumber\\
&&+\frac{b^\mathbf{m}_J}{\gamma}[1+\beta(\alpha^{\mathbf{m}}_{\rm eff}+\alpha^{\mathbf{m}}_{\rm sp})]\dot{\mathbf{n}}'-\frac{b^\mathbf{m}_J}{\gamma}[1+\beta(\alpha^{\mathbf{m}}_{\rm eff}+\alpha^{\mathbf{m}}_{\rm sp})](\mathbf{n}\cdot\dot{\mathbf{n}}')\mathbf{n}
+\frac{-\dot{\tilde{\gamma}}}{\tilde{\gamma^2}}\dot{\mathbf{n}}
-\frac{b^{\mathbf{n}}_J}{\gamma}[\beta-(\alpha^{\mathbf{m}}_{\rm eff}+\alpha^{\mathbf{m}}_{\rm sp})]\mathbf{n}\times\dot{\mathbf{n}}'+\dot{H}^{\mathbf{n}}_{\rm sh}n_y\mathbf{n}\nonumber\\
&=&-\frac{1}{\tilde{\gamma}}\ddot{\mathbf{n}}+\frac{1}{\gamma}(\alpha^{\mathbf{n}}_{\rm eff}-\alpha^{\mathbf{n}}_{\rm sp})\mathbf{n}\times\ddot{\mathbf{n}}-\bar{a}(\alpha^{\mathbf{m}}_{\rm eff}-\alpha^{\mathbf{m}}_{\rm sp})\dot{\mathbf{n}}-H^{\mathbf{n}}_{\rm sh}n_y\dot{\mathbf{n}}+(\alpha^{\mathbf{m}}_{\rm eff}+\alpha^{\mathbf{m}}_{\rm sp})\dot{\mathbf{f}}_{\mathbf{n}}+(\dot{\alpha}^{\mathbf{m}}_{\rm eff}+\dot{\alpha}^{\mathbf{m}}_{\rm sp})\mathbf{f}_{\mathbf{n}}+(\alpha^{\mathbf{m}}_{\rm eff}+\alpha^{\mathbf{m}}_{\rm sp})\mathbf{n}\times(\mathbf{f}_{\mathbf{n}}\times\dot{\mathbf{n}})\nonumber\\
&&-\frac{1}{\gamma}(\dot{\alpha}^{\mathbf{m}}_{\rm eff}+\dot{\alpha}^{\mathbf{m}}_{\rm sp})\beta b^\mathbf{m}_J\mathbf{n}'-\frac{1}{\gamma}(\dot{\alpha}^{\mathbf{m}}_{\rm eff}+\dot{\alpha}^{\mathbf{m}}_{\rm sp})b^{\mathbf{n}}_J\mathbf{n}\times\mathbf{n}'+(\dot{\alpha}^{\mathbf{m}}_{\rm eff}+\dot{\alpha}^{\mathbf{m}}_{\rm sp})H^{\mathbf{m}}_{\rm sh}\mathbf{n}\times\mathbf{y}-\bar{a}\gamma(\mathbf{n}\cdot\mathbf{f}_{\mathbf{n}})\mathbf{n}
-\bar{a}b^{\mathbf{n}}_J\mathbf{n}\times\mathbf{n}'.
\end{eqnarray}
Equations (\ref{eq:mag})-(\ref{eq:eom_n}) describe the field- and current-induced dynamics of one-dimensional conducting chiral AFMs with spatially and temporally varying texture. 
The dissipative torques, which are proportional to $\alpha_{\rm eff}^{{\bf m}({\bf n})}$, 
incorporate CD effects. To our best knowledge, these terms have not been derived previously. Specifically, Eq.~(\ref{eq:eom_n}) reveals that CD causes a chirality-dependent renormalization of rest mass and/or moment of inertia of AF textures (terms $\propto \ddot{\bf n}$). 
Since the rest mass is coupled to the ability of the dynamic magnetic texture to internally store energy, CD is anticipated to influence the dynamic deformation of textures like skyrmions \cite{Buttner2015,Lee2025}. This effect is also expected to apply to DWs in chiral ferromagnets, where CD results in a chirality-dependent shift in the Walker breakdown field/current \cite{Schryer1974}, which determines the onset of DW deformation \cite{Akosa2016, Akosa2024,Lee2024}.

We consider the dynamics of an AF DW by plugging into Eq.~(\ref{eq:eom_n}) a Walker profile~\cite{Schryer1974} as ${\bf n} = (\cos\phi\sin\theta, \sin\phi\sin\theta, \cos\theta)$, where $\phi = \phi(t)$ and $\theta(x,t) = 2 \tan^{-1} [\exp (s[x - X(t)]/\lambda)]$, with $X(t)$ and $\phi(t)$ being the DW center and tilt angle, respectively, and $s = \pm 1$ indicates $n_z=\ \uparrow,\downarrow$ or $\downarrow,\uparrow$ at $x=\mp\infty$.
For simplicity, we ignore distortions caused by dynamic DW width $\lambda$, assuming a rigid profile during motion. We also disregard the effects of nonlinear driving forces, such that $\ddot{\bf n} = \ddot{X} \partial_X{\bf n} + \ddot{\phi}\partial_\phi{\bf n}$ \cite{Tveten2013}. The effective field for $\bf n$ is taken as
\begin{eqnarray}
\mathbf{f}_{\mathbf{n}}&=&\mathbf{H}_{\mathbf{n}}+\frac{2Aa^2_0}{\mu_0 M_{\rm s}}\mathbf{n}''+\frac{2K}{\mu_0 M_{\rm s}}n_z{\mathbf{z}}+\frac{2 D_{\rm I}}{\mu_0 M_{\rm s}}(n'_z{\mathbf{x}}-n'_x\mathbf{z})+\frac{2 D_{\rm B}}{\mu_0 M_{\rm s}}(n'_z{\mathbf{y}}-n'_y{\mathbf{z}}),
\end{eqnarray}
where $A$ is the intra-sublattice ferromagnetic exchange stiffness ($a_0$ is lattice constant), $K$ is the easy-axis anisotropy energy, and $D_{\rm I}$ and $D_{\rm B}$ are respectively the interfacial and bulk DMI constants. The DMI density is
\begin{eqnarray}
\mathcal{H}_{\rm DM}&=&\frac{D_{\rm I}}{\mu_0 M_{\rm s}}(n_z\partial_x n_x-n_x\partial_x n_z)+\frac{D_{\rm B}}{\mu_0 M_{\rm s}}(n_z\partial_x n_y-n_y\partial_x n_z)=\frac{s D}{\mu_0 M_{\rm s}}\cos(\phi-\phi_{\rm dm})~\text{sech}\Big(\frac{x-X(t)}{\lambda}\Big),\\
D&\equiv&\sqrt{D^2_{\rm I}+D^2_{\rm B}},\ \phi_{\rm dm}\equiv\tan^{-1}(D_{\rm B}/D_{\rm I}).
\end{eqnarray}
It is important to note that the sign $s=\pm 1$ is fixed by the relative angle between DW tilt angle $\phi$ and $\phi_{\rm dm}$. On the other hand, for the local averaged and N\'eel CD parts, we take the model as
\begin{eqnarray}
\alpha^{\mathbf{m}}_{\rm eff}(x,t)&=&\alpha_0+\frac{s\Lambda^{\mathbf{m}}_{ \rm cd}}{\lambda}\cos(\phi-\phi^{\mathbf{m}}_{\rm cd})~\text{sech}\Big(\frac{x-X(t)}{\lambda}\Big),\\
\alpha^{\mathbf{n}}_{\rm eff}(x,t)&=&\frac{s\Lambda^{\mathbf{n}}_{\rm cd}}{\lambda}\cos(\phi-\phi^{\mathbf{n}}_{\rm cd})~\text{sech}\Big(\frac{x-X(t)}{\lambda}\Big),
\end{eqnarray}
where $\Lambda^{\mathbf{m,n}}_{\rm cd}$ are the strengths of CDs, and $\phi^{\mathbf{m}}_{\rm cd}$ are the angles characterizing the ratio between Bloch- and N\'eel-type CDs, similar to $\phi_{\rm dm}$ for DMIs.
We derive the equation of motion by examining the dynamics perpendicular to ${\bf n}$, which involves projecting Eq.~(\ref{eq:eom_n}) onto $\partial_x{\bf n}$ and ${\bf n} \times \partial_x{\bf n}$, followed by integrating over $x$. 
For simplicity, we ignore the hard-axis anisotropies and only consider a staggered magnetic field applied in $\bf z$ direction (${\bf H}_{\bf n}=H^z_{\bf n}{\bf z}$), and take the DW width $\lambda=a_0\sqrt{A/K}$ which is exact for the static DW without DMI and driving forces. Define $\bar{\alpha}^{\mathbf{m,n}}_{\rm c}\equiv\pi\Lambda^{\mathbf{m,n}}_{\rm cd}/(4\lambda)$ and $H_{\rm D}=D/(\mu_0 M_{\rm s}\lambda)$, after tedious but straightforward calculations, we end up with following two equations,
\begin{eqnarray}
\mathcal{F}_{X}&=&\mathcal{M}_{XX}\ddot{X}+\mathcal{M}_{X\phi}\ddot{\phi}+\mathcal{N}_{XX}\dot{X}+\mathcal{N}_{X\phi}\dot{\phi},\ \mathcal{F}_{\phi}=\mathcal{M}_{\phi X}\ddot{X}+\mathcal{M}_{\phi\phi}\ddot{\phi}+\mathcal{N}_{\phi X}\dot{X}+\mathcal{N}_{\phi\phi}\dot{\phi},\label{EQ}\\
\mathcal{M}_{XX}&=&\frac{1}{\gamma\lambda}\Big[2s(1 + \alpha_0^2-(\alpha_{\rm sp}^{\mathbf{m}})^2)+\bar{\alpha}^{\mathbf{m}}_{\rm c} \cos(\phi-\phi^{\mathbf{m}}_{\rm cd})\Big(4\alpha_0 + \frac{64}{3\pi^2}s\bar{\alpha}^{\mathbf{m}}_{\rm c}\cos(\phi-\phi^{\mathbf{m}}_{\rm cd})\Big)\Big]=\frac{-s}{\lambda}\mathcal{M}_{\phi\phi},\nonumber\\
\mathcal{M}_{X\phi}&=&\frac{2}{\gamma}\Big[\alpha_{\rm sp}^{\mathbf{n}}-s\bar{\alpha}^{\mathbf{n}}_{\rm c} \cos(\phi-\phi^{\mathbf{n}}_{\rm cd})\Big]=s\lambda\mathcal{M}_{\phi X},\nonumber\\
\mathcal{N}_{XX}&=&\frac{2s\bar{a}}{\lambda}\Big(\alpha_0 - \alpha_{\rm sp}^{\mathbf{m}} + s\bar{\alpha}^{\mathbf{m}}_{\rm c} \cos(\phi-\phi^{\mathbf{m}}_{\rm cd})\Big)+ \frac{s\pi}{2\lambda} H_{\rm sh}^{\mathbf{n}}\sin\phi
+\frac{4}{\gamma\lambda}\Big(\alpha_0 + \frac{32}{3\pi^2}s\bar{\alpha}^{\mathbf{m}}_{\rm c}\cos(\phi-\phi^{\mathbf{m}}_{\rm cd}) \Big)\dot{\bar{\alpha}}^{\mathbf{m}}_{\rm c} \cos(\phi-\phi^{\mathbf{m}}_{\rm cd}),\nonumber\\
\mathcal{N}_{\phi \phi}&=&-s\lambda \mathcal{N}_{XX}+s\pi H_{\rm D}\cos(\phi-\phi_{\rm dm})\Big(\alpha_0 + \alpha_{\rm sp}^{\mathbf{m}}+\frac{32}{3\pi^2}s\bar{\alpha}^{\mathbf{m}}_{\rm c}  \cos(\phi-\phi^{\mathbf{m}}_{\rm cd})\Big)\nonumber\\
&-&\frac{32}{3\pi}H_{\rm D}\sin(\phi-\phi_{\rm dm})\bar{\alpha}^{\mathbf{m}}_{\rm c} \sin(\phi-\phi^{\mathbf{m}}_{\rm cd})+\frac{2b^{\mathbf{n}}_J}{\lambda\gamma}\bar{\alpha}^{\mathbf{m}}_{\rm c}\sin(\phi-\phi^{\mathbf{m}}_{\rm cd}),\nonumber\\
\mathcal{N}_{X\phi}&=&-\frac{2}{\gamma}s\dot{\bar{\alpha}}^{\mathbf{n}}_{\rm c} \cos(\phi-\phi^{\mathbf{n}}_{\rm cd})+\frac{\pi}{2}H^{\mathbf{m}}_{\rm sh}\sin\phi\Big(\alpha_0+\alpha_{\rm sp}^{\mathbf{m}}+\frac{32}{3\pi^2}s\bar{\alpha}^{\mathbf{m}}_{\rm c}\cos(\phi-\phi^{\mathbf{m}}_{\rm cd})\Big)\nonumber\\
&+&2\Big(H^z_{\mathbf{n}}+\frac{s\beta b^\mathbf{m}_J }{\lambda\gamma}+\frac{4}{\pi}H^{\mathbf{m}}_{\rm sh} \cos\phi\Big)s\bar{\alpha}^{\mathbf{m}}_{\rm c}\sin(\phi-\phi^{\mathbf{m}}_{\rm cd}),\nonumber\\
\mathcal{N}_{\phi X}&=&-\frac{2}{\gamma\lambda}\dot{\bar{\alpha}}^{\mathbf{n}}_{\rm c} \cos(\phi-\phi^{\mathbf{n}}_{\rm cd})+\frac{s\pi}{2\lambda}H^{\mathbf{m}}_{\rm sh}\sin\phi\Big(\alpha_0 + \alpha_{\rm sp}^{\mathbf{m}}+\frac{16}{3\pi^2}s\bar{\alpha}^{\mathbf{m}}_{\rm c}\cos(\phi-\phi^{\mathbf{m}}_{\rm cd})\Big),\nonumber\\
\mathcal{F}_{X}&=& \bar{a}\gamma\Big(2H^z_{\mathbf{n}}+\frac{2s\beta b^{\mathbf{m}}_J}{\lambda\gamma}+\pi H^{\mathbf{m}}_{\rm sh} \cos\phi\Big)
+\Big(2H^z_{\mathbf{n}}+\frac{2s\beta b^{\mathbf{m}}_J }{\lambda\gamma}+\frac{8}{\pi} H^{\mathbf{m}}_{\rm sh}\cos\phi\Big)s\dot{\bar{\alpha}}^{\mathbf{m}}_{c}\cos(\phi-\phi^{\mathbf{m}}_{\rm cd})\nonumber\\
&+&2\dot{H}^z_{\mathbf{n}}\Big(\alpha_0 + \alpha_{\rm sp}^{\mathbf{m}}+ s\bar{\alpha}^{\mathbf{m}}_{\rm c} \cos(\phi-\phi^{\mathbf{m}}_{\rm cd})\Big)+\frac{2s}{\lambda\gamma}\dot{b}^\mathbf{m}_J
+\pi\dot{H}^{\mathbf{m}}_{\rm sh}\cos\phi\Big( \alpha_0 + \alpha_{\rm sp}^{\mathbf{m}} +\frac{8}{\pi^2 }s\bar{\alpha}^{\mathbf{m}}_{\rm c} \cos(\phi - \phi^{\mathbf{m}}_{\rm cd})\Big)\nonumber\\
&+&\frac{2s\beta \dot{b}^\mathbf{m}_J}{\lambda\gamma}\Big(\alpha_0+\alpha^{\mathbf{m}}_{\rm sp}+s \bar{\alpha}^{\mathbf{m}}_{\rm c}\cos(\phi-\phi^{\mathbf{m}}_{\rm cd})\Big),\nonumber\\
\mathcal{F}_{\phi}&=&-s\pi \dot{H}_{\rm D}\sin(\phi-\phi_{\rm dm})\Big(\alpha_0+\alpha^{\mathbf{m}}_{\rm sp}+\frac{32}{3\pi^2}s\bar{\alpha}^{\mathbf{m}}_{\rm c}\cos(\phi-\phi^{\mathbf{m}}_{\rm cd})\Big)
-\frac{32}{3\pi}H_{\rm D}\sin(\phi-\phi_{\rm dm})\dot{\bar{\alpha}}^{\mathbf{m}}_{\rm c}\cos(\phi-\phi^{\mathbf{m}}_{\rm cd})\nonumber\\
&-&\Big(2\dot{H}^{z}_{\textbf{m}}+\pi\dot{H}^{\mathbf{n}}_{\rm sh}\cos\phi
+\frac{2s\dot{b}^{\mathbf{n}}_J}{\lambda\gamma}
\Big(\beta-[\alpha_0 + \alpha_{\rm sp}^{\mathbf{m}}+s\bar{\alpha}^{\mathbf{m}}_{\rm c} \cos(\phi-\phi^{\mathbf{m}}_{\rm cd})]\Big)\Big)\nonumber\\
&+&\frac{2s\bar{a}b^{\mathbf{n}}_J}{\lambda}-\bar{a}\gamma s\pi H_{\rm D}\sin(\phi-\phi_{\rm dm})+\frac{2b^{\mathbf{n}}_J }{\lambda\gamma}\dot{\bar{\alpha}}^{\mathbf{m}}_{c}\cos(\phi-\phi^{\mathbf{m}}_{\rm cd}).\nonumber
\end{eqnarray}
These equations are the general equations of motion governing the dynamics of DW collective modes $X(t)$ and $\phi(t)$ in AFMs with DMI, under all the possible sublattice inequivalences of STT, SOT, spin pumping effect, CD, and the time derivatives of these driving forces. The dimensions of parameters are as follows: $\mu_0 M_{\rm s}\sim$ [T], $\mathcal{F}_\phi\sim$ [T/s], $H_{\rm D}\sim$ [T], $\bar{a}\sim$ [T], $H^{\mathbf{m,n}}_{\rm sh}\sim$ [T], $\gamma\sim$ [1/sT]. 
As a demonstration of the effects by sublattice-inequivalent CDs, we take $s=1$, ignore the applied magnetic fields ($H^z_{\bf m,n}=0$), spin pumping effect ($\alpha_{\rm sp}^{\bf m,n}=0$), and time derivatives of electric current and DMI, but keep the time oscillation of CD due to an applied ac gate voltage, then the these equations become Eqs.~(4)--(5) in the main text.

\section{Perturbative calculation of DW velocity and tilt angle}
In this section we show the details of the perturbative treatment of Eqs.~(4)--(5) in the main text. As outlined in the beginning of the Results section in the main text, we consider the case with Rashba SOI, ignore spin pumping effect, and take sublattice-equivalent STT and SOT. Assuming $\bar{\alpha}_{c}^{\mathbf{m,n}}=\bar{\alpha}_{c,0}^{\mathbf{m,n}}\sin(\Omega t)$, the $v$ and $\phi$ are expanded as in Eq.~(6) in the main text. After a perturbative expansion up to the first order of $\sin(\Omega t)$ and $\cos(\Omega t)$ in Eq.~(5) in main text, using the expressions of $\sin\phi$ and $\cos\phi$ as
\begin{eqnarray}
\sin\phi&\approx& \sin\phi_0+\cos\phi_0[\phi_s\sin(\Omega t)+\phi_c\cos(\Omega t)],\\
\cos\phi&\approx&\cos\phi_0-\sin\phi_0[\phi_s\sin(\Omega t)+\phi_c\cos(\Omega t)],
\end{eqnarray}
we get two equations,
\begin{eqnarray}
\mathcal{F}_{X}&=&\mathcal{M}_{XX}\ddot{X}+\mathcal{M}_{X\phi}\ddot{\phi}+\mathcal{N}_{XX}\dot{X}+\mathcal{N}_{X\phi}\dot{\phi}\Rightarrow 0=c_1+c_2\cos(\Omega t)+c_3\sin(\Omega t),\\
\mathcal{F}_{\phi}&=&\mathcal{M}_{\phi X}\ddot{X}+\mathcal{M}_{\phi\phi}\ddot{\phi}+\mathcal{N}_{\phi X}\dot{X}+\mathcal{N}_{\phi\phi}\dot{\phi}\Rightarrow 0=c_4+c_5\cos(\Omega t)+c_6\sin(\Omega t),
\end{eqnarray}
with coefficients $c_1, ..., c_6$ that can be straightforwardly calculated by using, e.g., Mathematica. It can be shown that $c_j$ only contains linear terms of the unknown variables in the main text, $v_0, v_s, v_c$,  $\phi_s, \phi_c$, and $c_j$ depends on $\sin\phi_0,\cos\phi_0$.
Due to the linear independences between constant, $\cos(\Omega t)$, and $\sin(\Omega t)$, at any time $t$, we have six equations as $0=c_j$ with $j=1, ..., 6$, which can give us the solutions of the six unknown variables. From $0=c_1,0=c_4$, one can get two equations,
\begin{eqnarray}
v_0&=&\frac{\beta b_J}{\alpha_0}+\frac{\pi\gamma H_{\rm sh}\lambda}{2\alpha_0}\cos\phi_0,\label{A}\\
0&=&\Big[\pi\bar{a}\gamma H_{\rm D}+\frac{\pi\alpha_0 H_{\rm sh}v_0}{2\lambda}\Big]\sin\phi_0.\label{D}
\end{eqnarray}
In Eq.~(\ref{D}), one solution is $\phi_0=0$ or $\pi$, and the other solution can be found by substituting $v_0$ in Eq.~(\ref{A}) into the bracket of Eq.~(\ref{D}) to get
\begin{eqnarray}
\cos\phi_0=-\frac{4\bar{a}H_{\rm D}}{\pi H_{\rm sh}^2}-\frac{2\beta b_J}{\pi\gamma H_{\rm sh}\lambda}\approx 3300,
\end{eqnarray}
by using our model parameters in the main text. This is not allowed since $|\cos\phi_0|\le 1$. Therefore, we get $\phi=0$ or $\pi$, which is determined by the initial DW tilt angle. When choosing the initial state with $\phi(0)=0$, the terminal $\phi_0=0$ is more probable when there is no strong phase switching effects, thus we get Eq.~(7) in the main text.

By setting $\phi_0=0$, the equations $0=c_j$ for $j=2,3,5,6$ have exact solutions as 
\begin{eqnarray}
v_s&=&-\frac{\bar{\alpha}^{\bf m}_{\rm c,0}}{2\pi[\alpha^2_0\gamma^2\bar{a}^2+(1+\alpha_0^2)^2\Omega^2]}\Big[\pi\gamma^2\bar{a}^2(2\beta b_J+\pi\gamma H_{\rm sh}\lambda)+2(1+\alpha_0^2)\Big(\pi\beta b_J+(\pi^2-4)\gamma H_{\rm sh}\lambda\Big)\Omega^2\Big],\\
v_c&=&\frac{\bar{\alpha}^{\bf m}_{\rm c,0}\gamma\bar{a}\Omega}{2\pi\alpha_0[\alpha^2_0\gamma^2\bar{a}^2+(1+\alpha_0^2)^2\Omega^2]}\Big[2\pi\beta b_J+\Big(\pi^2+(8-\pi^2)\alpha^2_0\Big)\gamma H_{\rm sh}\lambda\Big],\\
\phi_s&=&-\Big(16 \bar{\alpha}^{\bf n}_{\rm c,0}\gamma\lambda \Omega^2(2 \bar{a} - \pi H_{\rm D}) (2 \beta b_J+ \pi \gamma H_{\rm sh}\lambda)\Big)\Big/G,\\
\phi_c&=&\Big[4\bar{\alpha}^{\bf n}_{\rm c,0} \Omega (2 \beta b_J + \pi \gamma H_{\rm sh} \lambda)
\Big(\pi \gamma [2\beta b_J H_{\rm sh} + (4 \bar{a} H_{\rm D} + \pi H_{\rm sh}^2) \gamma \lambda] + 8 (1 + \alpha_0^2) \lambda \Omega^2\Big)\Big]\Big/(\alpha_0 G),\\
G&\equiv&\pi^2 \gamma^2 \Big(2 \beta b_J H_{\rm sh} + (4 \bar{a}H_{\rm D} + \pi H_{\rm sh}^2)\gamma\lambda\Big)^2 + 64 (1 + \alpha_0^2)^2 \lambda^2 \Omega^4\nonumber\\
&+&16 \gamma \lambda\Omega^2 \Big(2 \pi \beta b_J H_{\rm sh}(1 + \alpha_0^2) + \gamma \lambda [4 \pi \bar{a} H_{\rm D} + 4\alpha^2_0 \bar{a}^2 + \pi^2 (\alpha^2_0H_{\rm D}^2+(1+\alpha^2_0)H_{\rm sh}^2)] \Big)
\end{eqnarray}
By plugging these solutions into $\delta v=\sqrt{v^2_s+v^2_c}$, $\delta \phi=\sqrt{\phi^2_s+\phi^2_c}$, and expanding to the leading order of $\alpha_0$, after some tedious algebra we can get the expressions of Eq.~(8) in the main text.
\section{Case with oscillating DMI and CD}
As stated in the main text, when an ac gate voltage is applied to induce an oscillation of the strength of RSOI in time, in general it will lead to both the oscillations of DMI and CD simultaneously. 
To model this realistic situation, we consider the DMI field as
\begin{eqnarray}
H_{\rm D}(t)=H_{\rm D0}+\delta H_{\rm D}\sin(\Omega t).\label{DMI}
\end{eqnarray}
Substituting this DMI field into Eq.~(\ref{EQ}), retaining the terms proportional to $\delta H_{\rm D}\alpha_0$ but ignoring those proportional to $\delta H_{\rm D}\bar{\alpha}_c^{\bf m,n}$ for simplicity, we can solve Eq.~(\ref{EQ}) by a similar leading-order perturbative expansion to get
\begin{eqnarray}
&&v_0=\frac{\beta b_J}{\alpha_0}+\frac{\pi\gamma H_{\rm sh}\lambda}{2\alpha_0}\cos\phi_0,\  \phi_0=0,\\
&&\delta v\equiv\sqrt{v_s^2+v_c^2}\approx\frac{\bar{a}\alpha^{\mathbf{m}}_{\rm c,0}\gamma v_0 }{\Omega},\label{amp}\\
&&\delta\phi\equiv\sqrt{\phi_s^2+\phi_c^2}
\approx\frac{4\alpha_{\rm c,0}^{\mathbf{n}}\Omega v_0}
{\pi\alpha_0\gamma H_{\rm sh} v_0
	+2\lambda(\pi\gamma^2\bar{a}H_{\rm D0}+2\Omega^2)}.
\end{eqnarray}
Comparing these solutions to Eqs.~(7)--(8) in the main text, we find the only difference is the replacement of $H_{\rm D}$ by $H_{\rm D0}$ in the denominator of $\delta\phi$. 
There is no appearance of $\delta H_{\rm D}$ in these approximate solutions since it couples with the higher-order terms of $\sin(\Omega t)$ and $\cos(\Omega t)$.
To confirm these analytical results, Supplementary Fig.~\ref{SuppFig} shows the RK numerical results under oscillating DMI as Eq.~(\ref{DMI}) with $H_{\rm D0}=-1$~T and $\delta H_{\rm D}=0.5$~T (with all other parameters being the same as those used for Figs.~1~(a)--(b) in the main text) for DW velocity $\dot{X}$ in (a) and tilt angle $\phi$ in (b), which are in good agreement with the above equations as the black horizontal dashed lines from these equations give great estimate of the oscillation amplitudes of $\dot{X}$ and $\phi$, and $v_0$ matches the numerical averaged $\dot{X}$.
Moreover, Supplementary Figs.~\ref{SuppFig}~(a)--(b) are similar to Figs.~1~(a)--(b) in the main text, respectively. Therefore, our discussions about the behaviors of $\dot{X}$ and $\phi$ under various presences of $\bar{\alpha}_c^{\bf m,n}$ in the main text remain the same even in the realistic case with both oscillating DMI and CD induced by an ac gate voltage, provided the oscillating amplitude of DMI is small to support the perturbative treatment that gives the above solutions.
\begin{figure}[h]
	\centering
	\includegraphics[scale=0.55]{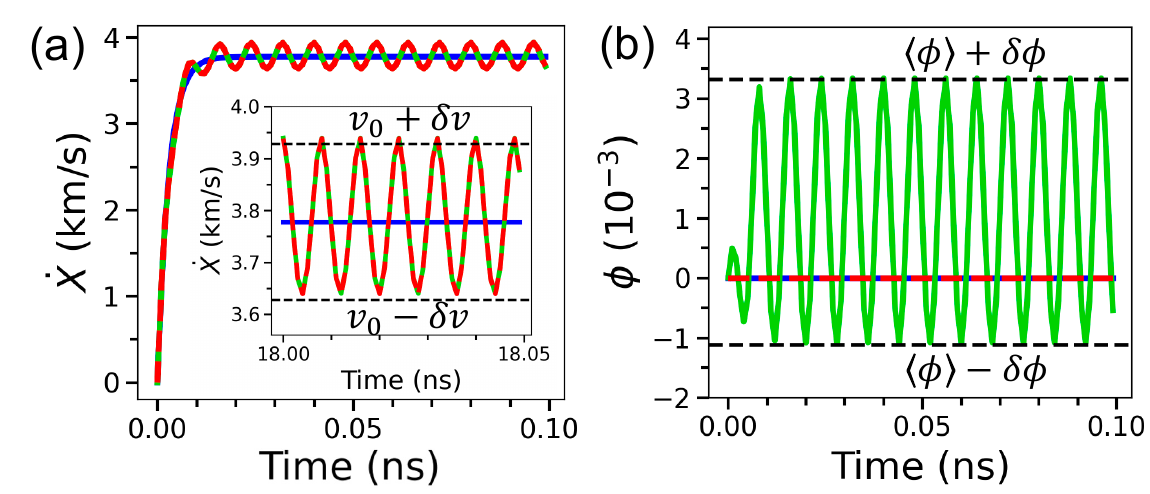}
	\renewcommand{\figurename}{Supplementary Fig.}\setcounter{figure}{0}
	\caption{Numerical results of DW (a) velocity and (b) tilt angle under SOT, STT, and both oscillating DMI and CD with frequency $\Omega$. The inset of (a) shows an enlarged view. The horizontal black dotted lines in the inset of (a) denote $v_0\pm \delta v$, while those in (b) denote $\langle\phi\rangle\pm\delta\phi$.}
	\label{SuppFig}
\end{figure}

\end{document}